\documentclass[aps,prb,twocolumn,groupedaddress]{revtex4}

\usepackage{graphicx}

\begin{document}

\title{Fluctuations of the Lyapunov exponent in 2D disordered systems}

\author{Keith Slevin}
\email[]{slevin@phys.sci.osaka-u.ac.jp}
\author{Yoichi Asada}
\affiliation{Department of Physics, Graduate School of Science,
Osaka University, Machikaneyama 1-1, Toyonaka-city, 560-0043 Osaka,
Japan}

\author{Lev I. Deych}
\affiliation{Department of Physics, Queens College of CUNY, Flushing, NY 11367, USA}

\date{\today}

\begin{abstract}
We report a numerical investigation of the fluctuations
of the Lyapunov exponent of a two dimensional non-interacting disordered system.
While the ratio of the mean to the variance of the Lyapunov exponent is not constant,
as it is in one dimension, its variation is consistent with the single parameter scaling hypothesis.
\end{abstract}

\maketitle

\section{Introduction}
%%%%%%%%%%%%%%%%%%%%%%%%%%%%%%%%%%%%%%%%%%%%%%%%%

The single parameter scaling (SPS) hypothesis\cite{wegner76,abrahams79} is the foundation of our
understanding of Anderson localization in disordered systems.
According to the SPS hypothesis Anderson localization phenomena are governed by
a single parameter: the ratio of the system size to the localization length $\xi$.
When applied to the zero temperature conductance $g$ of disordered mesoscopic system
the hypothesis implies that the probability distribution $p(g)$ of the conductance
obeys\cite{shapiro87}
\begin{equation}
p(g) \simeq F(g;L/\xi).
\label{eq:SPSPofG}
\end{equation}
Here $g$ is in units of $2e^2/h$ and $L$ is the system size. The SPS hypothesis has been
applied to other physically interesting quantities including the
localization length\cite{mackinnonPRL1981,mackinnon83} and Lyapunov exponent spectra\cite{slevin01b}
of quasi-1D systems, as well as the
energy level statistics of disordered systems.\cite{shklovskii93}
The probability distribution of all these quantities should have a form similar to 
Eq.(1) if SPS holds. 

Our understanding of scaling is most complete for one-dimensional (1D) systems.
There are two properties of 1D systems that distinguish them from higher dimensional systems.
First, their electronic eigenstates are, with very few exceptions, always localized.
Second, the localization length $\xi$ is comparable to the mean free path so that
there is no diffusive regime.
It has been shown that, in 1D, the cumulants of $\ln g$
all scale linearly with length $L$.\cite{roberts92}
It follows that $p(g)$ is log-normal when $L\gg \xi$.\cite{slevin90}
The log-normal distribution is determined by two parameters, the
mean and variance of $\ln g$.
Consistency with the SPS hypothesis requires that they
be related.
For weak disorder a perturbative analysis of the 1D Anderson model
reveals\cite{roberts92}
\begin{equation}
\sigma_{\ln g}^2
\equiv
\left< \left( \ln g - \left<\ln g \right> \right)^2 \right>
=
2\left<-\ln g \right>.
\label{eq:SPS1Dg}
\end{equation}
Angular brackets mean an average over disorder.
For the 1D Anderson model, only weak disorder is relevant since
for strong disorder $\xi$ is comparable to the lattice spacing.

Equation (\ref{eq:SPS1Dg}) holds for many models.
The precise conditions for its validity is\cite{deych98,deych00,deych01a,deych01b}
\begin{equation}\label{eq:SPScondition}
    \xi\gg\ell_s
\end{equation}
where $\ell_s$ is a length scale that is related to the integrated
density of states.
SPS is violated at the boundaries of the original spectrum of the
system and for fluctuation states arising due to disorder in the initial band gaps.
A violation\cite{roberts92,schomerus03} of SPS at the band center of the Anderson model
was shown in Ref.\onlinecite{deych03} to arise for similar reasons.

Single parameter scaling of the conductance distribution
(Eq.(\ref{eq:SPSPofG})) has also been verified numerically in the
three-dimensional (3D) Anderson model close to the critical point
of the Anderson transition.\cite{slevin01,slevin03} The region of
validity of the scaling in 3D, however, is not known. One can
imagine, however, that an inequality similar to Eq.
(\ref{eq:SPScondition}) may be applicable in this case as well.

The situation in two
dimensional (2D) systems is currently very controversial.
According to Ref.\onlinecite{abrahams79} all states in 2D are
localized. At the same time there are a large number of
experiments, in which an apparent metal-insulator transition has been
observed. (For a recent review see Ref.\onlinecite{abrahams2001}). The
physical meaning of these observation is not yet understood, despite
a debate that has already lasted a decade. The validity, or otherwise, of
SPS in 2D is, therefore, an important issue.
Even for single particle models this issue
has not yet been fully resolved.
For instance, careful numerical analyses\cite{mackinnonPRL1981,schreiberJPhys1992} of
the 2D Anderson model
showed excellent agreement
with SPS.
While other studies\cite{azbelPRB1982,kavehJPhys1985,kavehMott1985,pichardSarmaJPhys1985}
suggested the existence of power-law localized
states and two-parameter scaling.
Violations of SPS have also been reported in more recent papers.\cite{kantelhardt02,queiroz02}

The example of 1D systems demonstrates that
conclusive results regarding scaling properties can only
be obtained from studying the distribution functions of relevant
quantities.
Numerical studies\cite{chase1987,kramerPhilMag1992} in 2D show that
$\ln g$ is  normally distributed in the regime of strong localization.
It follows that single
parameter scaling must be manifest in a relation
between the average of $\ln g$, and its variance, similar to the
1D equation (\ref{eq:SPS1Dg}). However, attempts to verify
this relation did not
reach definite conclusions because of the small system sizes
simulated and an approach to Eq. (\ref{eq:SPS1Dg}) that was too naive.

The main objective of our paper is to perform a careful analysis of the
statistical properties of a 2D disordered system of non-interacting electrons, and verify that
they are consistent with SPS. The object of our calculations is
the finite length Lyapunov exponent (LE) for a 2D Anderson model
with diagonal disorder. (The definition of the LE and the meaning of
the qualification \emph{finite length} is given below.)
For a 2D $L\times L$ system with $L\gg \xi$ the mean of the LE
is equal to the inverse of the localization length $\xi$.

The distribution of conductance has been given special attention in the literature because 
it is directly accessible in experiments.
However, it should be understood that the conductance unavoidably reflects properties not 
only of the system in question, but also properties of the contacts used to measure it. 
The Lyapunov exponent, on the other hand, is an intrinsic property of the disordered system, 
containing information about spatial distribution of the wave functions, which are ultimately 
responsible for all other properties including conductance.

We find that the distribution function of the LE is approximately normal both when
$L\ll \xi$ and when $L\gg \xi$.
This contrasts with the conductance which exhibits  
not only very strong fluctuations but also a
significant change in the form of its distribution
between the diffusive and localized regimes.

We approach the question of scaling by clarifying the relation between the average and
variance of the LE that is implied by the SPS hypothesis in 2D, and checking whether
numerical data are consistent with it.
We find that the relation between the mean and variance is 
characterized by a single
parameter, namely the ratio of the system size to the localization length.
Thus, we provide convincing evidence that the SPS hypothesis is valid in 2D
disordered systems of non-interacting electrons.

\section{Model and method}

\subsection{The transfer matrix for the Anderson model }
We simulated the two dimensional Anderson model with Hamiltonian
\begin{equation}
H=\sum_i {\epsilon_ic_i^{\dag}c_i} - \sum_{\langle i,j\rangle}{c_i^{\dag}c_j}
\label{H}.
\end{equation}
The first summation is over all sites on an $L_t \times L$
square lattice i.e. a system of width $L_t$ and length $L$.
The second summation is over all pairs of
nearest neighbors.
We imposed periodic boundary conditions in the transverse direction and
used a ``box"
distribution of width $W$ for the site energies $\epsilon_i$
\begin{equation}
p(\epsilon_i)=\left\lbrace\begin{array}{ll}1/W&|\epsilon_i|\leq W/2\\
                        0 &|\epsilon_i|>W/2
                                  \end{array}\right.
\end{equation}

Lyapunov exponents arise when the time independent Schr\"odinger
equation is expressed as a product of random transfer
matrices.\cite{kramer93} We divide the system in the longitudinal
direction into $L$ layers. We form vectors $\Psi_n$ of length $L_t$
from the wavefunction amplitudes on each layer. For an
arbitrary energy $E$ we  derive from the Schr\"odinger
equation the transfer matrix equation
\begin{eqnarray}
\left(
\begin{array}{c}
\Psi_{n+1} \\
\Psi_{n}
\end{array}
\right) =M_n \left(
\begin{array}{c}
\Psi_{n} \\
\Psi_{n-1}
\end{array}
\right). \label{tmatdef}
\end{eqnarray}
The $2L_t \times 2L_t$ transfer matrix $M_n$ relates the wave function
amplitudes on layer $n$ and $n-1$ to those on layers $n$ and
$n+1$. For Eq. (\ref{H}), $\Psi_n$ and $M_n$ are real vectors and
matrices, respectively, and the transfer matrices are identically
and independently distributed random matrices.

\subsection{Definition of Lyapunov exponents}

We start with a $2L_t \times 2L_t$  orthogonal matrix $Q_0$.
We perform $L$ transfer matrix multiplications
and factor the result into a product of an orthgonal matrix $Q$, 
a diagonal matrix $D$ with positive elements,
and an upper triangular matrix $R$
with unit diagonal elements 
\begin{equation}
M_L \cdots M_1 Q_0 = Q D R.
\label{tmatproduct}
\end{equation}
We define $2L_t$ finite length LEs $\gamma_L^{(1)}\cdots\gamma_L^{(2L_t)}$ by
\begin{equation}
\gamma_L^{(n)}=\frac{1}{L}\ln D_n,
\label{eq:DefLE}
\end{equation}
Here $D_n$ is the $n$th diagonal element of $D$.
The finite length LEs are random variables that
fluctuate as we sample the random potential.
For fixed $L_t$, when $L\rightarrow \infty$ the LEs always tend to the
same limiting values
\begin{equation}
\lim_{ L  \rightarrow \infty } \gamma_L^{(n)} = \gamma^{(n)},
\label{eq:limit}
\end{equation}
for (nearly) all samplings of the distribution of transfer matrices and (nearly)
any choice of $Q_0$.\cite{crisanti93}

The $L_t^{\mathrm{th}}$ LE is the most physically significant: $\gamma^{(L_t)}$ is the inverse
of the localization
length $\lambda$ of an electron on an infinite quasi-1D system of
width $L_t$ described by (\ref{H})\cite{kramer93}
\begin{equation}
\gamma^{(L_t)}=\frac{1}{\lambda}.
\end{equation}
Therefore, in what follows we focus on $\gamma_L^{(L_t)}$, dropping the superscript
and referring to it as \emph{the} LE
\begin{equation}
\gamma_L^{(L_t)} \equiv \gamma_L.
\end{equation}

In numerical calculations, if only the first $m$ LEs are required,
it is sufficient to make $Q_0$ a $2L_t\times m$ real matrix with
orthonormal columns.
Depending on the value of $m$, this can save a considerable amount
of computer time.
The values  for the first $m$ LEs obtained in any particular calculation
are independent
of whether or not LEs with higher indices are also calculated.

To avoid numerical difficulties with the transfer matrix
multiplication Eq. (\ref{tmatproduct}), we performed additional
Gram-Schmidt orthogonalization after every 8 transfer matrix
multiplications.\cite{mackinnon83}

The definition, given in Eq. (\ref{eq:DefLE}), of the LE for \emph{finite length} that we have adopted here
is not the only reasonable one.
We compare our choice with an obvious alternative in Appendix \ref{AltDefAppendix}.

\subsection{Special considerations for systems of finite length}
\label{SpecialConsiderations}

In this paper we are concerned with the 
the distribution of $\gamma_L$ for finite length $L$
rather than with its asymptotic value as $L\rightarrow\infty$.
Therefore, we have to deal properly with effects related to the
finite value of $L$, effects that
were routinely considered unimportant in previous studies.

In the asymptotic limit $L\rightarrow\infty$, the value of $\gamma_L$ 
depends only on the distribution of the transfer matrix $p\left(M_n\right)$,
and is independent of the choice of the initial matrix $Q_0$.
For finite $L$, however, 
the distribution of $\gamma_L$ depends on $Q_0$ and $L$, in addition to $p\left(M_n\right)$.
The dependence on $Q_0$ would, if not dealt with, introduce an arbitrary element to
our analysis that is undesirable.

To remove the dependence of the distribution of $\gamma_L$
on $Q_0$ we used the
following observation to our advantage.
For (almost) any $Q_0$, the distribution of the matrix $Q$
approaches an $L$ independent stationary distribution $p_s\left(Q\right)$ as $L$ increases.
The form of $p_s\left(Q\right)$ depends only on $p\left(M_n\right)$.
By sampling $Q_0$ from $p_s\left(Q\right)$ 
we obtain a distribution for $\gamma_L$ that depends only on $L$ and 
$p\left(M_n\right)$.

To generate matrices with the required stationary distribution $p_s\left(Q\right)$, we took
an arbitrary set of
orthonormal vectors, performed $N_r$ transfer matrix multiplications
and factored the result according to Eq. (\ref{tmatproduct}).
To determine how large $N_r$ should be to get a good approximation
to $p_s\left(Q\right)$, we checked whether or not
the Kolmogorov-Smirnov test could distinguish between the
distributions of $\gamma_L$ for $L=1$ obtained with different
$N_r$.
The test showed that once $N_r>100$, the distribution of the LE
becomes independent of $N_r$. Below we set $N_r=1000$.

\section{Results}

Since our interest in this paper is in the distribution of
the LE in 2D systems we set the width $L_t$ and
length $L$ of the system equal
\begin{equation}
    L_t=L
\end{equation}
i.e. in the remainder of the paper
we consider only 2D $L\times L$ systems.

\subsection{Distribution of the LE}

\begin{figure}
\includegraphics[width=\linewidth]{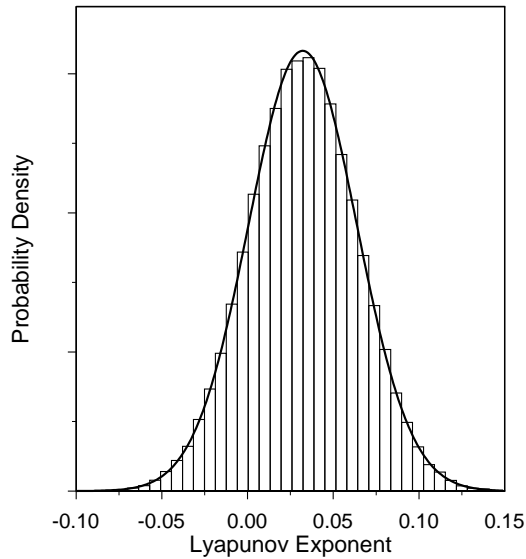}
\caption{\label{F1}The distribution of the LE for a 2D $L \times L$ system with
$E=1$, $W=5$ and $L=32$.
The line is a normal distribution with mean and variance equal to that
of the numerical data.
The numerical data have mean 0.032, standard deviation 0.032 and skewness -0.0031. 
The number of samples is 65,523.}
\end{figure}

\begin{figure}
\includegraphics[width=\linewidth]{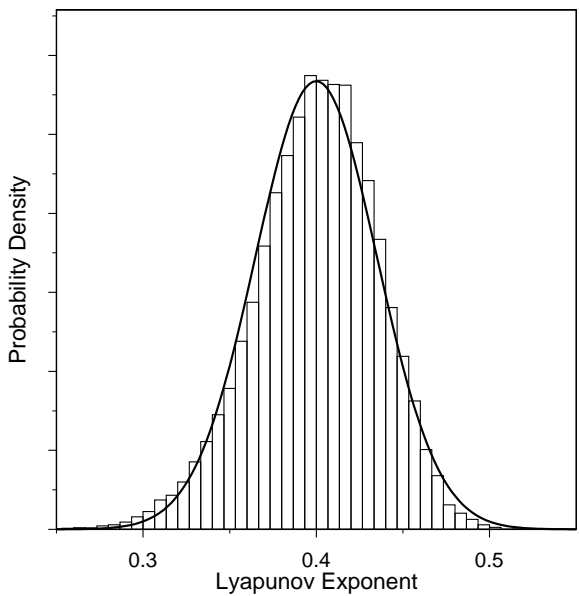}
\caption{\label{F2}The distribution of the LE for a 2D $L \times L$ system with
$E=1$, $W=14$ and $L=128$.
The line is a normal distribution with mean and variance equal to that
of the numerical data.
The numerical data have mean 0.40, standard deviation 0.035 and skewness -0.33. 
The number of samples is 40,000.}
\end{figure}

We simulated systems with Fermi energy $E=1$, disorder $5\le
W\le 14$ and a range of systems sizes between $L=16$ and $L=512$.
The distribution of the LE for two particular cases are shown in
Fig. \ref{F1} and Fig. \ref{F2}.
These are representative of the parameter range we studied.
Figure \ref{F1} corresponds to the situation $L\ll \xi$, while 
Fig. \ref{F2} corresponds to the situation $L\gg \xi$.
In the figure captions we give the values of the mean, variance and
skewness for the numerical data, as well as the number of samples simulated.

The skewness is a measure of the symmetry of the distribution.
Distributions
that are symmetrical about their mean, such as the normal distribution,
have a skewness equal to zero.
According to Ref. \onlinecite{bulmer}, a distribution whose
skewness has absolute value greater than unity is considered highly skew.
A distribution whose skewness has absolute value less than one half
is considered fairly symmetrical.
For data sampled from a normal distribution, the skewness is expected to
be distributed around zero with a standard deviation of $\sqrt{15/N_s}$ where
$N_s$ is the number of samples.\cite{numrep}

For the data in Fig. \ref{F1}, the difference of the skewness from zero is not statistically
significant. This is consistent with the LE having a normal distribution.

For the data in Fig. \ref{F2},
the difference of the skewness from zero is statistically significant.
What is the physical significance of this deviation? 
Normally we would
expect the scaling hypothesis to apply only when the localization length $\xi$
is much longer than microscopic length scales such as the mean free path,
lattice constant etc.
Here these are approximately unity, so this condition corresponds to
$\xi\gg 1$.
This condition is satisfied for the data in Fig. \ref{F1}, where 
$\xi\simeq180$ (see Table \ref{table:MeanLE}),
but not for the data in
Fig. \ref{F2}, where $\xi\simeq2.5$.
Therefore, we think that the deviation from the normal distribution seen in Fig. \ref{F2} is not
significant in the context of our study.

In our opinion, the normal distribution is a reasonable approximation to the
observed distribution for the range of $L/\xi$ in our simulations.
In what follows, we concentrate our attention on the mean and
variance of the LE and their scaling.

It is also important to bear in mind when looking at Figs. \ref{F1} and \ref{F2}
that the scaling hypothesis is expected to apply to the bulk of the distribution not its tails,
i.e. to typical states not necessarily to very rare states.
Hence, we use a linear scale for the probability density axis and not a logarithmic scale,
which would unduly emphasize the tails of the distribution.

\subsection{The scaling of the mean LE}

According to the SPS hypothesis the scaling of both the mean and variance of the LE
should be governed by the same length scale, the localization length $\xi$.
A quantitative test of scaling involves checking the consistency of
the disorder dependence of $\xi$ obtained independently
from the scaling of the mean and variance of the LE.
In this section we deal with the scaling of the mean LE.

\begin{figure}
\includegraphics[width=\linewidth]{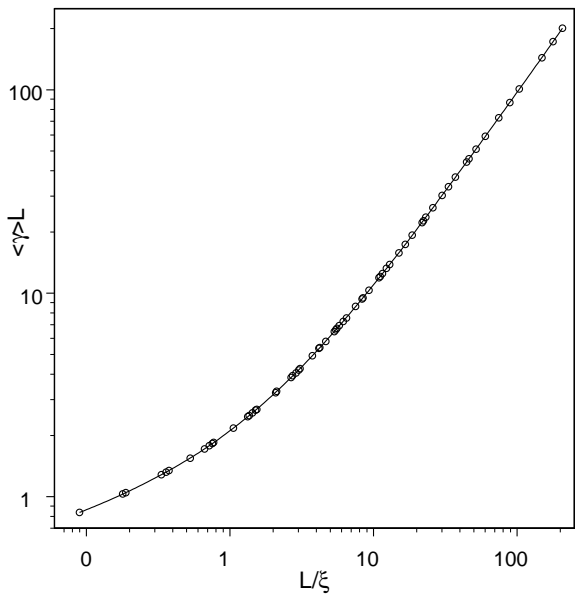}
\caption{\label{F3}
The finite size scaling fit (line) to the data (circles) for the mean of the
Lyapunov exponent.
The precision of the numerical data is $0.25\% $.}
\end{figure}

For the scaling analysis of the mean LE we estimated $\left<\gamma_L\right>$ to
a precision of $0.25\%$ for system sizes $L=16, 32, 64,
128, 256$ and $512$.
For $W\le 11$ the maximum system size was reduced to $256$.
We determined $\xi\equiv \xi\left(W\right)$ by fitting the variation of the
mean LE with $L$ and $W$ to the SPS law
\begin{equation}
\left<\gamma_L \right>L=F\left(\frac{L}{\xi}\right).
\end{equation}
When $L\gg \xi$ we suppose that the mean of the LE will tend to the inverse
of the 2D localization length $\xi$ i.e.
\begin{equation}
    \lim_{L\rightarrow\infty} \left\langle \gamma_{L}\right\rangle = \frac{1}{\xi}
\end{equation}
This is equivalent to
\begin{equation}
F\left(x \right) \rightarrow x \ \ \ x\gg 1.
\label{largex}
\end{equation}
For numerical reasons we expressed the scaling function in the form
\begin{equation}
\log_{10} \left<\gamma_L \right>L=f \left( \log_{10} \frac{L}{\xi} \right),
\end{equation}
and used a spline to interpolate the function $f$.
The values of $f$, at the values of $L/\xi$ in Table \ref{table:MeanLE}, were fitting parameters.
To ensure the spline interpolation reproduces Eq. (\ref{largex}), we fixed the value of $f$ at $L/\xi=1000$.
The corresponding value of $F$ is given in parenthesis in Table \ref{table:MeanLE}.
The remaining fitting parameters were
the localization lengths for each disorder.
Finally, we used the shape preserving Akima spline to avoid unphysical oscillations of $f$.
We summarize the results in Tables \ref{table:MeanLE} and \ref{table:fit} and in Figure \ref{F3}.
We can see from this figure that the data for different values of disorder $W$ and system size
$L$ fall on a common scaling curve when expressed as a function of $L/\xi$.
Moreover, for large $L$ we observe the expected linear dependence, with slope equal to the inverse localization length.

\begin{table}
\caption{\label{table:MeanLE}
The 2D localization length and the scaling function $F$ determined from the scaling of
the mean LE.
The errors quoted are $95\%$ confidence intervals and are estimated
using the Monte Carlo method.\cite{numrep}}
\begin{ruledtabular}
\begin{tabular}{|c|c||c|c|}
$W$ & $\xi$ & $L/\xi$   & $F$           \\ \hline
$5$ & $178\pm2$ & $0.1$     & $0.86\pm.005$     \\
$5.5$   & $85\pm1$  & $0.5$     & $1.51\pm.01$      \\
$6$ & $48\pm.5$ & $1$       & $2.11\pm.02$      \\
$6.5$   & $30.2\pm.3$   & $2$       & $3.17\pm.02$  \\
$7$ & $20.8\pm.2$   & $5$       & $6.16\pm.05$  \\
$8$ & $11.7\pm.1$   & $10$      & $11.0\pm.1$   \\
$9$ & $7.69\pm.08$  & $20$      & $20.6\pm.2$   \\
$10$    & $5.54\pm.06$  & $1000$    & ($1000$)  \\
$11$    & $4.26\pm.04$  &   \\
$12$    & $3.44\pm.03$  &   \\
$13$    & $2.88\pm.03$  &   \\
$14$    & $2.47\pm.02$  &   \\
\end{tabular}
\end{ruledtabular}
\end{table}

\begin{table}
\caption{\label{table:fit}
Details of the finite size scaling fits:
the number of data $N_d$, the number of parameters $N_p$, the value
of $\chi^2$ for the best fit and the goodness of fit probability $Q$.}
\begin{ruledtabular}
\begin{tabular}{|c|cccc|}
Statistic and data range    & $N_p$ & $N_d$ & $\chi^2$  & $Q$   \\ \hline
$\left<\gamma \right>L$ &   &   &       &   \\
$16\le L\le 512$, $5\le W\le 14$    & $19$  & $63$  & $46.0$    & $0.4$ \\ \hline
$\Sigma$            &   &   &       &   \\
$64\le L\le 256$, $5\le W\le 12$    & $15$  & $30$  & $20.3$        & $0.2$ \\
\end{tabular}
\end{ruledtabular}
\end{table}

\subsection{Scaling of the fluctuations of the LE}

Taking into account that the dimension of $\gamma_L$ is $1/L$, and
that of its variance $\sigma^2$ is $1/L^2$, we can define a
dimensionless quantity $\Sigma$ by
\begin{equation}
\Sigma=\frac{\sigma^2 L}{\left<\gamma_L \right>},
\label{eq:SigmaDef}
\end{equation}
According to the SPS hypothesis the localization length
is the only relevant length in the system, so
$\Sigma$ should obey the SPS law
\begin{equation}
\Sigma= F_{\Sigma}\left(\frac{L}{\xi}\right). \label{eq:SPSLE}
\end{equation}
In 1D, the linear scaling of the cumulants of $\ln g$, and the
relation between the LE and $g$ described in the Appendix, allow us
to deduce from Eq. (\ref{eq:SPS1Dg}) the much more prescriptive statement
\begin{equation}
\Sigma=1. \label{eq:SPS1DLE}
\end{equation}
However, for a 2D $L \times L$ system the cumulants of $\ln g$ do not scale
linearly with $L$, except perhaps in the regime where $L\gg \xi$;
a regime which it is more difficult to reach in 2D than in 1D.
Therefore, we should not expect that $\Sigma$ be unity or even constant
in our calculations.
Confirmation of the SPS hypothesis in 2D
consists not in demonstrating that calculated values vary in accord with
Eq. (\ref{eq:SPS1DLE}) but rather in trying to establish Eq. (\ref{eq:SPSLE}).

\begin{figure}
\includegraphics[width=\linewidth]{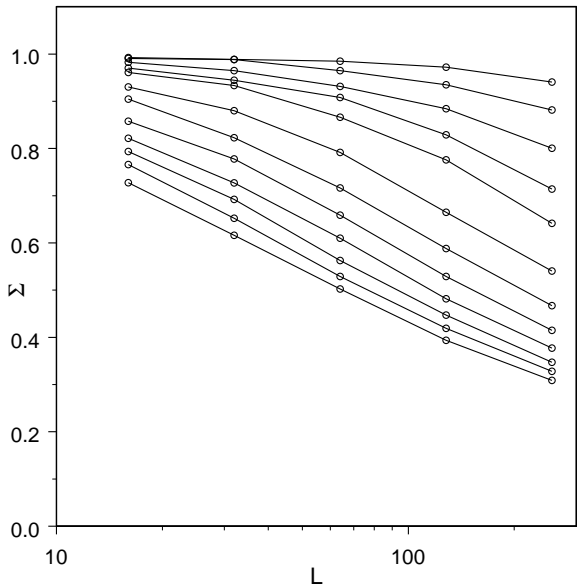}
\caption{\label{F4} $\Sigma$ versus system size. For each point
$160,000$ samples were simulated, corresponding to a
precision of approximately $0.4\% $. The lines, which connect points
corresponding to a common value of the disorder $W$, are a
guide to the eye only.}
\end{figure}

We have plotted the variation of $\Sigma$ with system size in
Figure \ref{F4}, where different lines correspond to different
values of disorder, $W$. These data were analyzed in an analogous
way to the mean LE. We expressed the SPS law Eq. (\ref{eq:SPSLE}) in
the form
\begin{equation}
\Sigma=f_{\Sigma}\left( \log_{10} \frac{L}{\xi} \right),
\end{equation}
and used an Akima spline interpolation of the function
$f_{\Sigma}$. The values of $f_{\Sigma}$, at the values of $L/\xi$ listed
in Table \ref{table:Sigma}, and the localization lengths for each
disorder were fitting parameters. To obtain a reasonable goodness
of fit ($>0.1$) we had to restrict the range of data considered to
$5\le W\le 12$ and $L\ge64$.
(There seem to be a more pronounced finite size correction in the data for the variance
than in the data for the mean LE.
Also, the breakdown of scaling when $\xi$ is comparable to the lattice spacing ($\xi\sim 1$)
seems to be evident sooner in the variance of the LE than in the mean LE.)

\begin{figure}
\includegraphics[width=\linewidth]{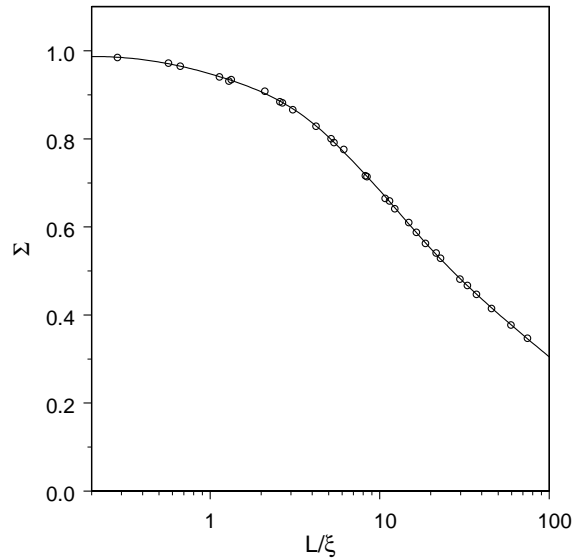}
\caption{\label{F5} A visual check that the numerical data for
$\Sigma$ defined in Eq. (\ref{eq:SigmaDef}) (points)
satisfy the SPS law Eq. (\ref{eq:SPSLE}) (line).}
\end{figure}

When fitting data for the mean LE, we were able to determine the
absolute value of $\xi$ with the aid of (\ref{largex}).
Unfortunately, no similar relation is available for
$f_{\Sigma}$ and so we cannot fix the absolute scale of $\xi$
by fitting data for $\Sigma$ alone.
Indeed, looking at Fig. \ref{F5} we can see that, if we translate both the fit and 
the data by the same amount parallel to the abscissa, we obtain an equally good fit. 
To avoid this
ambiguity, we set the value of the localization length for
$W=12$ to that found for the mean LE. We show the results in Table
\ref{table:Sigma}. Apart from an over estimate of $\xi$ for $W=5$
and $W=5.5$, the results are consistent with those for the mean
LE. In addition, in Figure $\ref{F5}$, we have plotted $\Sigma$
versus $L/\xi$, where $\xi$ is estimated from this fit. We
see that all the different curves of Fig.\ref{F4} collapse on to a single
curve, confirming the correctness of Eq. (\ref{eq:SPSLE}).
We conclude that the fluctuations
of the LE are consistent with the SPS hypothesis.

Looking at Fig. $\ref{F5}$, it is plausible that the function $\Sigma$ will tend to a finite
asymptotic value as $L/\xi\rightarrow\infty$.
If this does occur, the fluctuations of the LE in the
2D asymptotic limit ($L/\xi \rightarrow \infty$ with $L_t=L$)
decay as $1/\sqrt{L}$.
This is similar to the behavior in the quasi-1D limit ($L/\xi \rightarrow \infty$ with $L_t$ fixed)
where the fluctuations in the LE also decay as $1/\sqrt{L}$.
The only difference is that $\Sigma$ is always unity in quasi-1D, while  the
asymptotic value of $\Sigma$ is less than unity in 2D.

\begin{table}
\caption{\label{table:Sigma}
The 2D localization length and the scaling function $F$ determined from the scaling of
$\Sigma$.}
\begin{ruledtabular}
\begin{tabular}{|c|c||c|c|}
$W$ & $\xi$     & $L/\xi$   & $F_{\Sigma}$  \\ \hline
5   & $226\pm38$            & $0.1$     & $.98\pm.02$  \\
5.5 & $96\pm8$          & $1$       & $.948\pm.005$ \\
6   & $50\pm2$          & $5$       & $.804\pm.004$ \\
6.5 & $30.4\pm1$            & $10$      & $.683\pm.004$ \\
7   & $20.9\pm.6$           & $20$      & $.551\pm.003$ \\
8   & $11.9\pm.3$           & $50$      & $.403\pm.002$ \\
9   & $7.8\pm.2$            &       &       \\
10  & $5.6\pm.1$            &       &       \\
11  & $4.3\pm.1$            &       &       \\
12  & $(3.44)$          &       &       \\
\end{tabular}
\end{ruledtabular}
\end{table}

\section{Conclusion}

We have investigated numerically the scaling of the fluctuations
of the LE in the 2D Anderson model with diagonal disorder. We
found that the distribution of the LE is approximately normal both
when $L\ll \xi$ and $L\gg \xi$. We showed that the parameters of
the distribution, the mean LE and its variance
behave in accordance with the single parameter scaling hypothesis
for the energy value considered in our calculations. This value,
$E=1$, was chosen to lie far from the boundaries of the initial
spectrum, $E=\pm 4$, and from its center $E=0$ in order to avoid
anomalies related to the band edge behavior, which were found in
1D systems. We expect that the behavior
for any other value of the energy will be similar as long as it is not close to
an anomalous region.
We found that the manifestation of SPS in
numerical studies of 2D systems is different from that of 1D
systems. Instead of the simple relation between the mean and
variance of the LE given by Eq. (\ref{eq:SPS1DLE}) and valid for 1D, in 2D one
has to analyze the compliance of numerical data with the SPS relation Eq.
(\ref{eq:SPSLE}).

The fact that we verified SPS both when $L\ll \xi$ and $L\gg \xi$
is significant because it contradicts the conclusions of
Ref.\onlinecite{kantelhardt02} and Ref.\onlinecite{queiroz02}, where a behavior
inconsistent with SPS was found.
While a complete elucidation of the sources of this disagreement is
beyond the scope of this paper, we can suggest some possibilities
that might be worth pursuing in future work.
First, it is possible that the logarithmic increase of the localization
length seen in Ref.\onlinecite{kantelhardt02} might be reconcilable with the SPS
hypothesis, in much the same way as we have reconciled the system size and
disorder dependence of the ratio $\Sigma$ with SPS here.
Second, the authors of Ref.\onlinecite{kantelhardt02} and Ref.\onlinecite{queiroz02} analyzed
the spatial properties of eigenfunctions.
In 1D the relationship between lengths that characterize transport and
wavefunctions is well established.\cite{LGP}
In 2D there maybe aspects of this relationship that have not yet been
properly understood.
Third, in Ref.\onlinecite{kantelhardt02} and Ref.\onlinecite{queiroz02} wavefunctions
corresponding to $E=0$ were studied.
In 1D this is a special spectral point, at which SPS is violated.\cite{schomerus03,deych03}
It seems reasonable to
suggest that $E=0$ is also a special point for 2D
where SPS should not be expected.

\appendix
\section{Alternative definition of the LE for finite length. }
\label{AltDefAppendix}

The definition of the finite length LE we have used in this work is not the only
reasonable one.
In this appendix we will describe an alternative and compare with the definition
described in the main text of this paper.
Given a transfer matrix $M$
\begin{equation}
    M=\prod_{n=1}^{L} M_n,
\end{equation}
we can define a matrix $\Omega$ by
\begin{equation}
    \Omega=\ln M^{\dagger} M.
\end{equation}
The eigenvalue spectrum of $\Omega$ is composed of pairs of opposite sign
$\left\{+\nu^{\left(n\right)},-\nu^{\left(n\right)} : n=1\cdots L_t\right\}$.
From these eigenvalues we could define the LE in an alternative way as
\begin{equation}
\gamma_L^{(n)}=\frac{\nu^{\left(n\right)}}{2L}. \label{AltDef}
\end{equation}
In the limit that $L\rightarrow\infty$ at fixed $L_t$, the random variables
defined by Eq. (\ref{AltDef}) always tend to the same limiting values for (nearly) all samplings
of the distribution of transfer matrices.
These values are the same as those obtained with Eq. (\ref{eq:DefLE}) in the same limit.
For finite $L$ the values of Eq. (\ref{eq:DefLE}) and Eq. (\ref{AltDef}) are different.
We summaries the main characteristics of each definition below.

For the definition Eq. (\ref{eq:DefLE}) in the main text:
\begin{description}
  \item{P1} The LE are not the eigenvalues of a matrix. The indices of the LE refer to the order
  in which they are obtained from the Gram-Schmidt procedure.
  In general, this is \emph{not} in a strictly decreasing order.
    \item{P2} Though the sum of the all LEs is always zero, for finite $L$ and for a single sample,
    the LE do not occur in pairs of opposite sign. This symmetry is restored
    after  taking the limit $L\rightarrow\infty$ for a single sample,
    or after averaging over an ensemble of samples.
    For a single sample we have found that the symmetry also appears when $Q_0$ is sampled from
    the stationary distribution $p_s\left(Q_0\right)$ described in \ref{SpecialConsiderations}.
    \item{P3}\label{Lindep} For fixed $L_t$ and $Q_0$ sampled from $p_s\left(Q_0\right)$, the mean of the
    LEs are independent of $L$. (Note that in the main text we consider scaling with
    $L_t=L$, so this property is not applicable there.)
    \item{P4} The LE have a simple geometrical interpretation in terms of the exponential
    rate of increase of lengths, areas, volumes etc.
\end{description}

For the definition (\ref{AltDef}):
\begin{description}
    \item{P1} The LEs are related to the eigenvalues of a matrix and hence there is no
    prescribed ordering for them. It is conventional to put the LEs in decreasing order
    and the index in the definition (\ref{AltDef}) usually refers to this order.
    \item{P2} For all $L$, the LE occur in pairs of opposite sign. This exact symmetry is exhibited not
    just after averaging over an ensemble of samples but also by a single sample.
    \item{P3} In this definition there is no analogue of $Q_0$ and hence no analogue of property P3
    for Eq. (\ref{eq:DefLE}).
    \item{P4} There is no simple geometric interpretation except in the asymptotic limit.
\end{description}
SPS can be investigated using either definition. The quantity defined by Eq. (\ref{eq:DefLE})
has the advantage that its distribution is normal, while at the same time
retaining a straightforward relationship to the decay of the wavefunction in
the disordered system.

\section{Relation of LE to conductance in 1D.}
\label{LEtoGAppendix}
For a strictly 1D system whose length is much longer than the
localization length the transmission coefficient $t$ for the
transmission of electrons through the disordered sample
decays as
\begin{equation}
-\ln \left| t \right| = \ln D_1 + O(L^0) \ \  \left( L\gg \xi \right).
\end{equation}
The $O(L^0)$ term is a fluctuating term that depends on the nature of the leads
attached to the sample when defining the scattering problem.
Using the Landauer formula to relate the transmission and the conductance
we have
\begin{equation}
-\ln g  =  2 \gamma_L L + O(L^0) \ \ \left( L\gg \xi \right).
\end{equation}
From this we deduce that equation Eq. (\ref{eq:SPS1Dg}) is equivalent to
equation Eq. (\ref{eq:SPS1DLE}) when $L\gg \xi$.

\begin{acknowledgments}
Keith Slevin would like to thank Tomi Ohtsuki for suggesting scaling at a
fixed aspect ratio.
Lev Deych is grateful to Alex Lisyansky for numerous useful discussions of this work.
\end{acknowledgments}

\end{document}